%Paper: 9108011
%From: SEN%TIFRVAX.BITNET@cunyvm.cuny.edu
%Date: Wed, 21 Aug 91 19:49 IST

%THIS FILE NEEDS TO BE PROCESSED USING THE MACRO PHYZZX.TEX
%\overfullrule=0pt
%\scrollmode
\input phyzzx

{}~\hfill\vbox{\hbox{TIFR/TH/91-37}\hbox{August, 1991}}

\title{TWISTED BLACK $p$-BRANE SOLUTIONS IN STRING THEORY}

\author{Ashoke Sen\foot{e-mail address: SEN@TIFRVAX.BITNET}}

\address{Tata Institute of Fundamental Research, Homi Bhabha Road,
Bombay 400005, India}

\abstract

It has been shown that given a classical background in string theory
which is independent of $d$ of the space-time coordinates, we can generate
other classical backgrounds by $O(d)\otimes O(d)$ transformation on the
solution.
We study the effect of this transformation on the known
black $p$-brane solutions in string theory, and show how these
transformations produce new classical solutions labelled by extra continuous
parameters and containing background antisymmetric tensor field.

\NPrefs
\def\define#1#2\par{\def#1{\Ref#1{#2}\edef#1{\noexpand\refmark{#1}}}}
\def\con#1#2\noc{\let\?=\Ref\let\<=\refmark\let\Ref=\REFS
         \let\refmark=\undefined#1\let\Ref=\REFSCON#2
         \let\Ref=\?\let\refmark=\<\refsend}

\define\DVV
R. Dijkgraaf, E. Verlinde and H. Verlinde, preprint PUPT-1252,
IASSNS-HEP-91/22.

\define\WITTEN
E. Witten,     preprint IASSNS-HEP-91/12.

\define\MVE
K. Meissner and G. Veneziano, preprint CERN-TH-6138/91.

\define\VENEZIA
G. Veneziano, preprint CERN-TH-6077/91.

\define\MANDAL
G. Mandal, A.M. Sengupta and S.R. Wadia, preprint IASSNS-HEP-91/10.

\define\MMS
S. Mukherji, S. Mukhi and A. Sen, preprint TIFR/TH/91-28.

\define\VENEZIA
G. Veneziano preprint CERN-TH-6077/91.

\define\ROCEK
M. Rocek, K. Schoutens and A. Sevrin, preprint IASSNS-HEP-91/14.

\define\BARDACKI
K. Bardacki, M. Crescimannu, and E. Rabinovici, Nucl.
Phys. {\bf B344} (1990) 344.

\define\SCALE
A.Sen preprint IC/91/195 (TIFR-TH-91-35).

\define\CALLAN
C.G. Callan, D. Friedan, E. Martinec and M. Perry, Nucl. Phys. {\bf B262}
(1985) 593.

\define\EFR
S. Elitzur, A. Forge and E. Rabinovici, Preprint RI-143-90.

\define\BANE
I. Bars and D. Nemeschansky, Nucl. Phys. {\bf B348} (1991) 89.

\define\SCHWARZ
J. Schwarz, Phys. Rep. {\bf 89} (1982) 223.

\define\FT
E. Fradkin and A. Tseytlin, Phys. Lett. {\bf 158B} (1985) 316.

\define\LOVELACE
C. Lovelace, Phys. Lett. {\bf 135B} (1984) 75.

\define\SEN
A. Sen, Phys. Rev. {\bf D32} (1985) 2102; Phys. Rev. Lett. {\bf 55} (1985)
1846.

\define\LISTEIF
N. Ishibashi, M. Li and A.R. Steif, preprint UCSB-91-28.

\define\CHS
C. G. Callan, J. Harvey and A. Strominger, Nucl. Phys. {\bf B359} (1991)
611.

\define\HOST
G. Horowitz and A. Strominger, Nucl. Phys. {\bf B360} (1991) 197.

\define\GIDST
S. Giddings and A. Strominger, preprint UCSBTH-91-35.

\define\GHS
D. Garfinkle, G. Horowitz and A. Strominger, preprint UCSB-TH-90-66.

\define\DGHR
A. Dabholkar, G. Gibbons, J. Harvey and F. Ruiz, Nucl. Phys. {\bf B340}
(1990) 33.

\define\DUFFLU
M.J. Duff and J. Lu, preprint CTP-TAMU-81/90.

\define\CMP
C. G. Callan, R. C. Myers and M. Perry, Nucl. Phys. {\bf B311} (1988) 673.

\define\MYERS
R.C. Myers, Nucl. Phys. {\bf B289} (1987) 701.

\define\GIBBONS
G. Gibbons, Nucl. Phys. {\bf B207} (1982) 337; G. Gibbons and K. Maeda,
Nucl. Phys. {\bf B298} (1988) 741.

\define\VESA
H.J. de Vega and N. Sanchez, Nucl. Phys. {\bf B309} (1988) 552.

\define\MAZUR
P. Mazur, Gen. Rel. and Grav. {\bf 19} (1987) 1173.

\define\MYPE
R. C. Myers and M. Perry, Ann. Phys. {\bf 172} (1986) 304.

\define\VAFAB
R. Brandenberger and C. Vafa, Nucl. Phys. {\bf B316} (1989) 391.

\def\p{\partial}
\def\hy{\hat Y}
\def\ty{\tilde Y}
\def\odd{O(d)\otimes O(d)}

\def\r{\rangle}
\def\tg{\tilde G}
\def\hg{\hat G}
\def\tb{\tilde B}
\def\hb{\hat B}

\def\th{\tilde H}

\def\tr{\tilde R}
\def\tp{\chi}
\def\odo{O(d-1,1)\otimes O(d-1,1)}

\endpage

In a previous paper\SCALE\ we showed that in string theory, if we have an
exact classical solution which is independent of $d$ of the space-time
coordinates, then we can perform an $O(d)\otimes O(d)$ transformation on
the solution, which produces a new configuration of string field,
satisfying the classical equations of motion to all orders in the string
tension $\alpha'$.
This generalised the result found by Meissner and Veneziano\VENEZIA\MVE\
to leading order in $\alpha'$.
In this paper we shall study the effect of this $\odd$ transformation on
the known black $p$-brane solutions in string
theory, and obtain new classical solutions in string theory labelled by
extra continuous parameters.

We begin by recalling the general argument of ref.\SCALE\ and also by
giving a generalised version of the analysis of ref.\MVE.
In the language of string field theory, looking for solutions which are
independent of $d$ of the coordinates (say $Y^i$, $1\le i\le d$)
corresponds to looking for a string state $|\Psi\r$ carrying zero momentum
in these $d$ directions.
Restricting string states of this type gives us the reduced string field
theory action which governs the classical dynamics in this subspace.
This reduced action, in turn, is expressed in terms of correlation
functions of vertex operators carrying zero $Y^i$ momentum in the
appropriate conformal field theory.
In the part of the conformal field theory described by the free scalar
fields $Y^i$, the correlation functions factorise into the left and the
right part, and each part is separately invariant under the rotation group
$O(d)$ which acts on these $d$ coordinates.
Thus the reduced action has an $\odd$ symmetry, which implies that given a
classical solution of the string field theory equations of motion in the
subspace carrying zero $Y^i$ momemta, we can generate other solutions by
acting with this $O(d)\otimes O(d)$ transformation.
Of this the diagonal $O(d)$ subgroup simply corresponds to rotating the
solution in the $d$ dimensional space, the other generators of $\odd$
acting on the solution produces inequivalent solutions in general, since
$\odd$ is not a symmetry of the full action.
Although the above argument was given in the context of string field
theory, note that the argument is independent of the detailed form of
string field theory, and hence the final result is expected to hold for
fermionic string theories as well.
Note that if one of the coordinates $Y^m$ is time-like, the $\odd$
transformation gets replaced by $\odo$.

The low energy manifestation of this symmetry had been discovered in
ref.\MVE.
We shall briefly reproduce this analysis in a somewhat more general form
than the one in which it was discussed in ref.\MVE.
Let us consider the low energy effective action of string theory in $D$
space-time dimension.
This can be obtained either from the study of the $S$-matrix elements in
string theory (see ref.\SCHWARZ\ and references therein) or from the
calculation of the $\beta$-function of
the $\sigma$-model\con\LOVELACE\FT\SEN\CALLAN\noc, and is given by,
$$
S=-\int d^Dx\sqrt{\det G}e^{-\Phi}(\Lambda-R^{(D)}(G)+{1\over 12}
H_{\mu\nu\rho} H^{\mu\nu\rho}-G^{\mu\nu}\p_\mu\Phi\p_\nu\Phi)
\eqn\eone
$$
where $G_{\mu\nu}$, $B_{\mu\nu}$ and $\Phi$ denote the graviton, the
dilaton, and the antisymmetric tensor fields respectively,
$H_{\mu\nu\rho}=\p_\mu B_{\nu\rho}$ + cyclic permutations, $R^{(D)}$
denotes the $D$ dimensional Ricci scalar, and $\Lambda$ is the
cosmological constant equal to $(D-26)/3$ for bosonic string and
$(D-10)/2$ for fermionic string.
(For simplicity we have set the other background massless fields, which
appear in fermionic string theories, to zero.)
Let us now split the coordinates $X^\mu$ into two sets $\hat Y^m$ and
$\tilde Y^\alpha$ ($1\le m\le d$, $1\le \alpha\le D-d$) and consider
backgrounds independent of $\hat Y^m$.
Let us further concentrate on backgrounds where
$G_{m\alpha}=B_{m\alpha}=0$,
i.e. to backgrounds of the form $G=\pmatrix{\hg_{mn} & 0 \cr 0 &
\tg_{\alpha\beta}}$, $B=\pmatrix{\hb_{mn} & 0 \cr 0 & \tb_{\alpha\beta}}$.
In this case, after an integration by parts, the action \eone\ can be
shown to take the form:
$$\eqalign{
- \int d^d\hy \int & d^{D-d} \ty \sqrt{\det \tg}
e^{-\tp}\Big[\Lambda-\tg^{\alpha\beta}\tilde
\p_\alpha\tp\tilde \p_\beta\tp -{1\over 8}\tg^{\alpha\beta} Tr(\tilde
\p_\alpha
M L \tilde \p_\beta M L)\cr
&-\tr^{(D-d)}(\tg)+{1\over
12}\th_{\alpha\beta\gamma}\th^{\alpha\beta\gamma}\Big]\cr
}
\eqn\etwo
$$
where,
$$
L=\pmatrix{0 & \bf 1\cr \bf 1 & 0}
\eqn\etwoa
$$
$$
\tp=\Phi-\ln\sqrt{\det \hg}
\eqn\ethree
$$
and,
$$
M=\pmatrix{\hg^{-1} & -\hg^{-1}\hb\cr \hb\hg^{-1} & \hg-\hb \hg^{-1}\hb}
\eqn\efour
$$
If the coordinates $\hy^m$ are all of Euclidean signature, this action is
invariant under  an $O(d)\otimes O(d)$ transformation on $\hg$, $\hb$ and
$\Phi$, given by,
$$
M\to {1\over 4}\pmatrix{S+R & R-S\cr R-S & S+R} M \pmatrix{S^T+R^T &
S^T-R^T\cr S^T-R^T & S^T+R^T}
\eqn\efive
$$
$$
\tp\to\tp,~~~\tg_{\alpha\beta}\to\tg_{\alpha\beta},~~~
\tb_{\alpha\beta}\to \tb_{\alpha\beta}
\eqn\esix
$$
where $S$ and $R$ are $O(d)$ rotation matrices, and $S^T$, $R^T$ denote
the transpose of the matrices $S$, $R$.
In fact, the action is invariant under a general $O(d,d)$ transformation
which leaves the matrix $L$ invariant\MVE, but the members of the $O(d,d)$
algebra outside the $\odd$ algebra can be shown to generate pure gauge
deformations\SCALE\ if the coordinates $\hy^m$ are non-compact.
The transformations given in eqs.\efive, \esix\ were shown to agree with
the $\odd$ transformation on the string fields to linearised order.
Corrections to the action given in eq.\eone\ including higher derivative
terms are expected to change the transformation laws given in eqs.\efive,
\esix, but the existence of a modified transformation is guaranteed by the
string field theory argument given before.

Since in the above analysis we have explicitly set $G_{m\alpha}$ and
$B_{m\alpha}$ to zero, we must make sure that the equations of motion
obtained by varying the action with respect to these fields is also $\odd$
invariant.
In this case it is easy to see that for the backgrounds considered here
these equations of motion are satisfied identically, hence their $\odd$
invariance is obvious.

The above symmetry gets modified to $\odo$ when one of the coordinates
$\hy^m$ (say $\hy^1$) is time-like.
In this case the transformation laws \efour\ get modified to,
$$
M\to {1\over 4}\pmatrix{\eta(S+R)\eta & \eta(R-S)\cr (R-S)\eta & S+R} M
\pmatrix{\eta(S^T+R^T)\eta & \eta(R^T-S^T)\cr (R^T-S^T)\eta & (S^T+R^T)}
\eqn\efivena
$$
where $\eta=$diag($-1,1,\ldots 1)$, and $S$ and $R$ are $O(d-1,1)$
matrices satisfying,
$$
S\eta S^T=\eta,~~~~R\eta R^T=\eta
\eqn\efivenc
$$

In all the examples we shall consider, one of the coordinates $\hy^m$ will
be time-like, and hence the relevant group will be $\odo$.
We shall now examine the transformation laws of various fields under the
$\odo$ group in some detail.
In component form the transformed fields $\hg'_{ij}$, $\hb'_{ij}$ and
$\Phi'$ are given by,
$$\eqalign{
(\hg'^{-1})_{ij}=& {1\over 4}\big[\eta(S+R)\eta \hg^{-1}\eta (S^T+R^T)\eta
+\eta(R-S)(\hg-\hb\hg^{-1}\hb)(R^T-S^T)\eta\cr
&-\eta(S+R)\eta \hg^{-1}\hb (R^T-S^T)\eta+\eta(R-S)\hb\hg^{-1}\eta
(S^T+R^T)\eta\big]_{ij}\cr
\hb'_{ij}=&{1\over 4}\Big(\big[(R-S)\eta\hg^{-1}\eta(S^T+R^T)\eta+
(S+R)(\hg-\hb\hg^{-1}\hb)(R^T-S^T)\eta\cr
&+(S+R)\hb\hg^{-1}\eta(S^T+R^T)\eta-(R-S)\eta\hg^{-1}\hb(R^T-S^T)
\eta\big]\hg'\Big)_{ij}\cr
\Phi'=&\Phi-{1\over 2}\ln\det\hg+{1\over 2}\ln\det\hg'\cr
}
\eqn\efivea
$$
The transformation laws of $\hg$ and $\hb$ may be expressed in a compact
form by defining the matrix:
$$
C\equiv
\hg-\eta\hg^{-1}\eta-\hb\hg^{-1}\hb+\hb\hg^{-1}\eta+\eta\hg^{-1}\hb
\eqn\efiveb
$$
The transformation law of $C$ then takes a simple form:
$$
C'=SCR^T
\eqn\efivec
$$

Let us now turn to specific examples.
We start from a simple solution of the equation of motion \eone\ of the
form\MANDAL\WITTEN:\foot{The $\odo$ transformation may also be applied to
the more general class of solutions discussed in ref.\MMS, but we shall
not discuss it here.}
$$\eqalign{
ds^2=&-\tanh^2{Qr\over 2}dt^2+dr^2+\sum_{i=1}^{d-1}dX^idX^i\cr
B_{\mu\nu}=&0\cr
\Phi=&-\ln\cosh^2{Qr\over 2}+\Phi_0\cr
}
\eqn\eseven
$$
where $\Phi_0$ is an arbitrary parameter, and,
$$\eqalign{
Q=\sqrt{-\Lambda}=&\sqrt{25-d\over 3}~~{\rm for~bosonic~string}\cr
=&\sqrt{9-d\over 2}~~{\rm for~fermionic~string}\cr
}
\eqn\eeight
$$
Since the solution is independent of $d-1$ space-like coordinates, it can
be called a $d-1$ brane solution in the language of ref.\HOST.
Note that {\it a priori} we cannot ignore the contribution from the higher
derivative terms in the action since the scale of spatial variation of the
solution is set by $Q$ which is of order 1.
This problem may be avoided if instead of just taking $d-1$ scalar
(super)-fields $X^i$, we take $d-1$ scalar (super)-fields together with a
(super)-conformal field theory of central charge $c$ (${3\over 2}c$).
In this case $Q$ will be given by,
$$\eqalign{
Q=&\sqrt{25-d-c\over 3}~~{\rm for~bosonic~string}\cr
=&\sqrt{9-d-c\over 2}~~{\rm for~fermionic~string}\cr
}
\eqn\enine
$$
and we may consider a situation where $d+c\simeq 25$ ($d+c\simeq 9$) so
that $Q$ is small.
In this case the effect of higher derivative terms in the effective action
will be small, at least away from any singularity.

The solution given in eq.\eseven\ has the property that it is independent
of the coordinates $t$ and $X^i$.
Thus we can make an $\odo$ transformation on the fields so that the new
configuration will also be a solution of the equations of motion.
For the background defined in eq.\eseven\ the tensor $C_{ij}$ defined in
eq.\efiveb\ takes the form:
$$
C_{ij}=-\delta_{i1}\delta_{j1}(\tanh^2{Qr\over 2}-\coth^2{Qr\over 2})
\eqn\eten
$$
Thus from eq.\efivec\ we get,
$$
C'_{ij}=-S_{i1}R_{j1}(\tanh^2{Qr\over 2}-\coth^2{Qr\over 2})
\eqn\eeleven
$$
In order to study the set of inequivalent classical solutions, we note
that under a Lorentz transformation, $C'\to MC'M^T$ where $M$ is an
$O(d-1,1)$ matrix.
This corresponds to changing $S$ to $MS$ and $R$ to $MR$.
Thus,
$$
S_{i1}\to M_{ik}S_{k1},~~~~~~R_{i1}\to M_{ik}R_{k1}
\eqn\etwelve
$$
Both the vectors $S_{i1}$ and $R_{i1}$ are normalised to $-1$ with respect
to the metric $\eta$ since they form the first columns of $O(d-1,1)$
matrices.
Thus by choosing an appropriate $M$ we may bring the vectors $S_{i1}$ and
$R_{j1}$ into the form:
$$\eqalign{
S_{11}=\cosh\theta,~~S_{21}=-\sinh\theta,~~S_{i1}=0~~{\rm for}~i\ge 3\cr
R_{11}=\cosh\theta,~~R_{21}=\sinh\theta,~~R_{i1}=0~~{\rm for}~i\ge 3\cr
}
\eqn\ethirteen
$$
For this choice of $S$ and $R$ eq.\efivea\ gives the following form of the
transformed fields $\hg'$, $\hb'$ and $\Phi'$:
$$\eqalign{
\hg'=&
\pmatrix{-{\sinh^2(Qr/2)\over\cosh^2(Qr/2)+\sinh^2\theta}
&0&0&\cdots
&0\cr
0 & {\cosh^2(Qr/2)\over\cosh^2(Qr/2)+\sinh^2\theta} &0 &\cdots &0\cr
0&0&1&\cdots &0\cr
\cdot&\cdot&\cdot&\cdots&\cdot\cr
\cdot&\cdot&\cdot&\cdots&\cdot\cr
0&0&0&\cdots&1\cr}
\cr
\hb'=&{\cosh\theta\sinh\theta\over\cosh^2(Qr/2)+\sinh^2\theta}
\pmatrix{0&1&0&\cdots&0\cr
-1&0&0&\cdots&0\cr
0&0&0&\cdots&0\cr
\cdot&\cdot&\cdot&\cdots&\cdot\cr
\cdot&\cdot&\cdot&\cdots&\cdot\cr
0&0&0&\cdots&0\cr
}
\cr
\Phi'=&-\ln(\cosh^2(Qr/2)+\sinh^2\theta)+\Phi_0\cr
}
\eqn\efourteen
$$
Thus the full metric now takes the form:
$$\eqalign{
ds^2=&-{\sinh^2(Qr/2)\over \cosh^2(Qr/2)+\sinh^2\theta}dt^2+dr^2
+{\cosh^2(Qr/2)\over\cosh^2(Qr/2)+\sinh^2\theta}(dX^1)^2\cr
&+\sum_{i=2}^{d-1}dX^i dX^i\cr
}
\eqn\efifteen
$$
and has a coordinate singularity at $r=0$.
This singularity may be removed by choosing a new coordinate system:
$$
u=\sinh(Qr/2)e^{Qt/(2\cosh\theta)},~~~ v=\sinh(Qr/2)e^{-Qt/(2\cosh\theta)}
\eqn\esixteen
$$
In this coordinate system the metric takes the form:
$$\eqalign{
ds^2=&{2\over Q^2}dudv\bigg({\cosh^2\theta\over uv
+\cosh^2 \theta}+{1\over
uv+1}\bigg)\cr
&-{1\over Q^2}{\sinh^2\theta\over (uv+1)(uv+\cosh^2\theta)} (v^2du^2+
u^2dv^2)\cr
&+{uv+1\over
uv+\cosh^2\theta}(dX^1)^2+\sum_{i=2}^{d-1} dX^idX^i\cr
}
\eqn\eseventeen
$$
Thus we see that at $r=0$, i.e. at $u=0$ or $v=0$, the metric does not
have any singularity in this new coordinate system.
In fact, in this coordinate system the metric has finite non-zero
eigenvalues in the region $uv>-1$, $u$, $v$ finite.
In order to make sure that the solution is non-singular at $r=0$, we must
also check that the dilaton, as well as the field strength associated with
the
anti-symmetric tensor field are non-singular at $r=0$.
The dilaton given in eq.\efourteen\ is clearly non-singular at $r=0$.
The antisymmetric tensor field strength $H_{\mu\nu\rho}=(\p_\mu
B_{\nu\rho}$ + cyclic permutations) takes the form:
$$
H_{rt1}=\p_rB_{t1}={-Q\cosh\theta\sinh\theta\cosh(Qr/2)\sinh(Qr/2)\over
(\cosh^2(Qr/2)+\sinh^2\theta)^2}
\eqn\eeighteen
$$
When transformed to the $u-v$ coordinate system, this becomes,
$$
H_{uv1}={2\over Q}{\cosh^2\theta\sinh\theta\over (uv+\cosh^2\theta)^2}
\eqn\enineteen
$$
which is clearly non-singular at $u=0$ or $v=0$.

So far we have discussed the part of $\odo$ transformation that is
connected to identity.
As was discussed in ref.\SCALE, the effect of the disconnected part of the
$\odd$ transformation on the solution is to expand the solution in powers
of $e^{Qr}$ and reverse the sign of all the odd powers of $e^{Qr}$.
This transforms the solution given in eq.\efourteen\ to,
$$\eqalign{
\hg'=&
\pmatrix{-{\cosh^2(Qr/2)\over\sinh^2(Qr/2)-\sinh^2\theta} &0&0
&\cdots&0 \cr
0&{\sinh^2(Qr/2)\over\sinh^2(Qr/2)-\sinh^2\theta}&0&\cdots&0\cr
0&0&1&\cdots &0\cr
\cdot&\cdot&\cdot&\cdots&\cdot\cr
\cdot&\cdot&\cdot&\cdots&\cdot\cr
0&0&0&\cdots&1\cr}
\cr
\hb'=&-{\cosh\theta\sinh\theta\over\sinh^2(Qr/2)-\sinh^2\theta}
\pmatrix{0&1&0&\cdots&0\cr
-1&0&0&\cdots&0\cr
0&0&0&\cdots&0\cr
\cdot&\cdot&\cdot&\cdots&\cdot\cr
\cdot&\cdot&\cdot&\cdots&\cdot\cr
0&0&0&\cdots&0\cr
}
\cr
\Phi'=&-\ln(\sinh^2(Qr/2)-\sinh^2\theta)+\Phi_0\cr
}
\eqn\etwenty
$$
Thus in this case the solution is singular at $r=2\theta/Q$.

As was shown by Witten\WITTEN\ (see also
refs.\con\EFR\BARDACKI\ROCEK\DVV\BANE\noc), for the solution given in
eq.\eseven, after modification due to higher order terms have  been taken
into account, the coordinates $r$ and $t$ together describe a solvable
conformal field theory based on the gauged SL(2,R) WZW theory, where a
non-compact subgroup of SL(2,R) is gauged.
The level $k$ of the WZW theory is related to $Q$ by the relation:
$$
2+3Q^2={3k\over k-2}-1
\eqn\etwentyone
$$
so that the $Q\to 0$ limit corresponds to $k\to\infty$.
It is thus natural to ask whether there are solvable conformal field
theories which correspond to the solution given in eq.\efourteen.
In fact, conformal field theories associated with Euclidean continuation
of these solutions have been found in ref.\LISTEIF\ by
gauging a linear combination of a compact U(1)
generator of SL(2,R) and the U(1) currents generated by $\p X^i$, $\bar\p
X^i$.
By taking independent linear combinations of the U(1) currents in the left
and the right sector one can get the Euclidean continuation of the
background given in eq.\efourteen\ (with $\theta$ replaced by $i\theta$).
This is not surprising, since the original $\odd$ symmetry was due to the
freedom of independent rotations in the left and the right sector.

Note that if we take some of the directions $X^i$ to be compact, then even
the diagonal $O(d-1,1)$ subgroup, acting on the solution, generates new
solutions, since this is no longer a symmetry of the full theory.
This has been exploited in ref.\LISTEIF\ to get new exact solutions of
string theory.
In the analysis of this paper we shall restrict ourselves to the
non-compact case.

We shall now consider a second example, where the solution, in general,
does not correspond to a solvable conformal field theory.
This is the five-brane solution of the low energy effective field theory
in ten dimensional superstring or heterotic string theory (or,
equivalently, the 21-brane solution in 26 dimensional bosonic string
theory) described in ref.\HOST.
(See also
refs.\con\CHS\GIDST\GHS\DGHR\DUFFLU\CMP\MYERS\GIBBONS\VESA\MAZUR\MYPE\noc
for related
work.)
In a particular coordinate system the solution takes the form\GIDST:
$$\eqalign{
ds^2=&-\tanh^2r dt^2+[M+\delta^2(\cosh^2r-{1\over 2})](dr^2+d\Omega_3^2)
+\sum_{i=1}^5 dX^idX^i\cr
\Phi=&\ln\bigg({M+\delta^2( \cosh^2r-{1\over 2})\over\delta^2 \cosh^2 r}
\bigg)+\Phi_0 \cr
H=& 2 Q_0\epsilon_3\cr
}
\eqn\etwentyfour
$$
where $M$ and $\Phi_0$ are independent continuous parameters,
$$
Q_0\equiv \sqrt {M^2-{1\over 4}\delta^4}
\eqn\etwentyfive
$$
is quantised, $d\Omega_3$ is the line element on the 3-sphere, and
$\epsilon_3$ is the volume form on the same 3-sphere.\foot{The dilaton
$\Phi$ and the field strength $H$ in our convention are related to that in
the convention of
ref.\HOST\ by a factor of 2.}
(For bosonic string theory the sum over $i$ runs from 1 to 21 instead of
from 1 to 5.)
Although the solution does not appear to be flat in the asymptotic region
$r\to\infty$, after a coordinate transformation:
$$
y=\delta\cosh r
\eqn\etwentysix
$$
one gets a metric which reduces to 10 dimensional Minkowski metric in the
region $r\to\infty$.

Let us note that the solution is independent of the coordinates $t$ and
$X^i$.
Thus, as before, we can construct new solutions through $\odo$
transformation (here $d=6$).
Counting of independent parameters labelling the transformed solution
proceeds exactly in the same way as before, and we
are left with one independent parameter $\theta$.
The new solution obtained in this way is given by:
$$\eqalign{
ds^2=&-{\sinh^2 r\over \cosh^2 r+\sinh^2\theta} dt^2+{\cosh^2r\over
\cosh^2r+\sinh^2\theta} (dX^1)^2\cr
&+(M+\delta^2(\cosh^2r-{1\over 2}))(dr^2+d\Omega_3^2)+\sum_{i=2}^5
dX^idX^i \cr
\Phi=& \ln{M+\delta^2(\cosh^2 r-{1\over 2})\over \delta^2(\cosh^2
r+\sinh^2\theta)} +\Phi_0\cr
H^{(int)}=& 2 Q_0\epsilon_3\cr
\hb=&{\cosh\theta\sinh\theta\over\cosh^2r+\sinh^2\theta}
\pmatrix{0&1&0&\cdots&0\cr
-1&0&0&\cdots&0\cr
0&0&0&\cdots&0\cr
\cdot&\cdot&\cdot&\cdots&\cdot\cr
\cdot&\cdot&\cdot&\cdots&\cdot\cr
0&0&0&\cdots&0\cr
}
\cr
}
\eqn\etwentyseven
$$
where $H^{(int)}$ denotes the field strength associated with the
antisymmetric tensor field in the internal 3-sphere, and $\hb$ denotes the
components of the antisymmetric tensor field in the six dimensional space
spanned by the coordinates $t$ and $X^i$.
As before, it is easy to see that the point $r=0$ represents a coordinate
singularity, and the field strength associated with the antisymmetric
tensor field is regular at $r=0$.
Thus we see that using the $\odo$ transformation we can generate a
solution of the low energy effective field theory equations of motion
characterised by three continuous parameters $\Phi_0$, $M$ and $\theta$,
and one discrete parameter $Q_0$.

In conclusion, we have demonstrated in this paper that the $\odd$ ($\odo$)
symmetry may be used effectively to generate new classical solutions in
string theory from the known ones.
Although we have considered only a few examples, it is clear that this
transformation can be applied on other known classical solutions as well.
It will be interesting to study the the physical consequences of
continuous parameter family of solutions in string theory implied by the
$\odd$ symmetry, particularly on the cosmology of the early universe in
string theory\VAFAB.

Finally, note that for suitable backgrounds, the $\odo$ symmetry of the
reduced action may be extended to $O(d-1,1)\otimes O(d+15,1)$ symmetry in
the case of heterotic string theory, due to the possibility of including
the 16 internal coordinates in the rotation.
This rotation, in general, will produce a charged black hole\GHS\ from an
uncharged one.
Thus besides producing new solutions, our method also opens up the
possibility of relating different known solutions in string theory by the
twisting procedure.

\refout
\end